\documentclass[a4paper]{jpconf}
\usepackage{amsmath}
\usepackage{amssymb}
\usepackage{graphicx}
\usepackage{cite}
\begin{document}
\title{Probing the linearly polarized gluons in unpolarized proton with heavy-quark pair production}

\author{N. Ya. Ivanov$^{1,3}$, A.V. Efremov$^2$, O. V. Teryaev$^{1,2}$}
\address{$^1$ Veksler and Baldin Laboratory of High Energy Physics, JINR, 141980 Dubna, Russia}
%\author{A.V. Efremov}
\address{$^2$ Bogoliubov Laboratory of Theoretical Physics, JINR, 141980 Dubna, Russia}
%\ead{efremov@theor.jinr.ru}
%\author{O. V. Teryaev}
\address{$^3$ Yerevan Physics Institute, Alikhanian Br.~2, 0036 Yerevan, Armenia}
%\ead{teryaev@theor.jinr.ru}
\ead{nikiv@yerphi.am}

\begin{abstract}
We consider the azimuthal $\cos \varphi$ and $\cos 2\varphi$ distributions and the Callan-Gross 
ratio $R={\rm d}\sigma_L/{\rm d}\sigma_T$ in heavy-quark pair electroproduction, 
$lN\rightarrow l^{\prime}Q\bar{Q}X$, as probes of linearly polarized gluons in unpolarized nucleons. Our analysis shows that the azimuthal asymmetries and Callan-Gross ratio are predicted to be large and very sensitive to the contribution of the gluonic counterpart of the Boer-Mulders function, $h_{1}^{\perp g}$, describing the linear polarization of gluons inside unpolarized nucleon. In particular, the maximum values of the azimuthal distributions vary from 0 to 1 depending on $h_{1}^{\perp g}$. We conclude that future measurements of these quantities at the proposed EIC and LHeC colliders could clarify in details the proton spin decomposition puzzle.
\end{abstract}

\section{Introduction}
Transverse momentum dependent (TMD) distributions of the transversely polarized quarks, so-called Boer-Mulders function $h_{1}^{\perp q}$ \cite{Boer-Mulders}, and linearly polarized gluons, $h_{1}^{\perp g}$ \cite{Mulders_2001}, in an unpolarized nucleon play especial role in studies of the spin-orbit couplings of partons and understanding of the proton spin decomposition. It is also well-known that the heavy flavor production in lepton-proton DIS offers direct access to the gluon distributions in the proton. In Refs.~\cite{Boer_HQ_1,Boer_HQ_2,Boer_HQ_3}, the azimuthal correlations in heavy-quark pair production in unpolarized DIS have been proposed as probes of the $h_{1}^{\perp g}$ density. The complete angular structure of the $Q\bar{Q}$ production cross section has been obtained in terms of seven azimuthal modulations at LO in QCD. As shown in Ref.~\cite{we_compass}, only two of these modulations are really independent; they can be chosen as the $\cos \varphi$ and $\cos 2\varphi$ distributions, where $\varphi$ is the heavy quark (or anti-quark) azimuthal angle.

In Ref.~\cite{we_TMD}, the $\cos \varphi$ and $\cos 2\varphi$ asymmetries in heavy flavor pair leptoproduction have been studied as probes of the linearly polarized gluons in unpolarized nucleon. Analogous analysis for the Callan-Gross ratio $R={\rm d}\sigma_L/{\rm d}\sigma_T$ was performed  in Ref.~\cite{we_TMD_CG}. It is shown that the azimuthal distributions and ratio $R$ are predicted to be large but very sensitive to the gluonic counterpart of the Boer-Mulders function,  $h_{1}^{\perp g}$. In particular, the maximum value of the Callan-Gross ratio $R$ varies from 0 to $\frac{Q^2}{4m^2}$ depending on the polarized gluon density $h_{1}^{\perp g}$.

In this talk, we discuss in details the sensitivity of the pQCD predictions for the azimuthal  asymmetries and ratio $R={\rm d}\sigma_L/{\rm d}\sigma_T$ to the contribution of linearly polarized gluons inside unpolarized proton. In particular, we consider the $p_{\perp}$- and $z$\,-  distributions of these quantities for the case of charm and bottom production at energies of the proposed EIC \cite{EIC} and LHeC \cite{LHeC2} colliders. 

\section{Linearly polarized gluons in unpolarized proton}
\subsection{Production cross section}
We discuss the contribution of the linearly polarized gluons to the reaction
\begin{equation} \label{1}
l(\ell )+N(P)\rightarrow l^{\prime}(\ell -q)+Q(p_{Q})+\bar{Q}(p_{\bar{Q}})+X(p_{X}) 
\end{equation}
with unpolarized initial states. To describe the contributions of TMD densities, the sum and difference of the transverse heavy quark momenta are used in the plane perpendicular to the direction of the target and the exchanged photon: $\vec{K}_{\perp}=\frac{1}{2}(\vec{p}_{Q\perp}-\vec{p}_{\bar{Q}\perp})$, $\vec{q}_{T}=\vec{p}_{Q\perp}+\vec{p}_{\bar{Q}\perp}$. Following  \cite{Boer_HQ_1,Boer_HQ_2,Boer_HQ_3}, we use the approximation when $\vec{q}_{T}^2\ll \vec{K}_{\perp}^2$, $\vec{p}^2_{Q\perp}\simeq \vec{p}^2_{\bar{Q}\perp}\simeq\vec{K}_{\perp}^2$, and the outgoing heavy quark and anti-quark are almost back-to-back in the transverse plane.

The contribution of the photon-gluon fusion mechanism to the reaction (\ref{1}) has the following factorized form:
\begin{equation} \label{2}
{\rm d}\sigma\propto L(\ell,q)\otimes \Phi_g(\zeta, k_{T})\otimes \left| H_{\gamma^*g\rightarrow Q\bar{Q}X} (q,k_{g},p_{Q},p_{\bar{Q}})\right|^2, 
\end{equation}
where $L^{\alpha\beta}(\ell,q)$ is the leptonic tensor and $H_{\gamma^*g\rightarrow Q\bar{Q}X}(q,k_{g},p_{Q},p_{\bar{Q}})$ is the amplitude for the hard partonic subprocess. The symbol $\otimes$ stands for appropriate convolutions. Information about parton densities in unpolarized nucleon is formally encoded in  corresponding TMD parton correlators. In particular, the gluon correlator is usually parameterized  as \cite{Mulders_2001}
\begin{equation} \label{3}
\Phi_g^{\mu\nu}(\zeta, k_{T})\propto - g_T^{\mu\nu}f_{1}^{g}\big(\zeta,\vec{k}_{T}^2\big)+\left(g_T^{\mu\nu}-2\frac{k_T^\mu k_T^\nu}{k_T^2}\right)\frac{\vec{k}_{T}^2}{2m^2_N}h_{1}^{\perp g}\big(\zeta,\vec{k}_{T}^2\big), 
\end{equation}   
where $m_N$ is the nucleon mass. In Eq.~(\ref{3}), the tensor $-g_T^{\mu\nu}$ is (up to a factor) the density matrix of unpolarized gluons. The TMD distribution $h_{1}^{\perp g}\big(\zeta,\vec{k}_{T}^2\big)$ describes  the contribution of linearly polarized gluons. The degree of their linear polarization is determined by the quantity $r=\frac{\vec{k}_{T}^2 h_{1}^{\perp g}}{2m^2_N f_{1}^{g}}$. In particular, the gluons are completely polarized along the $\vec{k}_{T}$ direction at $r=1$, $\vec{k}_g=\zeta \vec{P}+\vec{k}_{T}$. 

The LO predictions for the azimuth-dependent cross section of the reaction (\ref{1}) can be written as \cite{we_TMD}
\begin{align} 
&\frac{{\rm d}^{6}\sigma}{{\rm d}y{\rm d}x{\rm d}z{\rm d}\vec{K}_{\perp}^2{\rm d}\vec{q}_{T}^2{\rm d}\varphi}=\frac{e_{Q}^{2}\alpha_{em}^2\alpha_{s}}{8\bar{S}^2 y^3 x}\frac{f_{1}^{g}(\zeta,\vec{q}_{T}^2)\hat{B}_2}{\zeta z\,(1-z)}\Bigg\{\left[1+(1-y)^2 \right]\left(1-2r \frac{\hat{B}^h_2}{\hat{B}_2}\right)-y^2\frac{\hat{B}_L}{\hat{B}_2}\left(1-2r \frac{\hat{B}^h_L}{\hat{B}_L}\right) \nonumber \\
&+2(1-y)\frac{\hat{B}_A}{\hat{B}_2}\left(1-2r \frac{\hat{B}^h_A}{\hat{B}_A}\right)\cos2\varphi +(2-y)\sqrt{1-y}\frac{\hat{B}_I}{\hat{B}_2}\left(1-2r \frac{\hat{B}^h_I}{\hat{B}_I}\right)\cos\varphi\Bigg\}, \label{4}
\end{align}
where $e_Q$ is the heavy quark charge; $x$, $y$ and $Q^2$ are the usual Bjorken variables; $\bar{S}=2\, \ell\cdot P$, $z=\frac{p_{Q}\cdot P}{q\cdot P}$, $\zeta=\frac{q\cdot p_{Q}}{q\cdot P}$  and $\varphi$ is the heavy quark azimuth.

The coefficients $\hat{B}_i$ $(i=2,L,A,I)$ originate from the contribution of unpolarized gluons,  while the quantities $\hat{B}^h_i$ are associated with the function $h_{1}^{\perp g}$. The LO results for $\hat{B}_i$ and $\hat{B}^h_i$ $(i=2,L,A,I)$ are presented in Ref.~\cite{we_TMD}.

\subsection{Azimuthal asymmetries}
\begin{figure}[h]
\begin{minipage}{18.2pc}
\includegraphics[width=18.2pc]{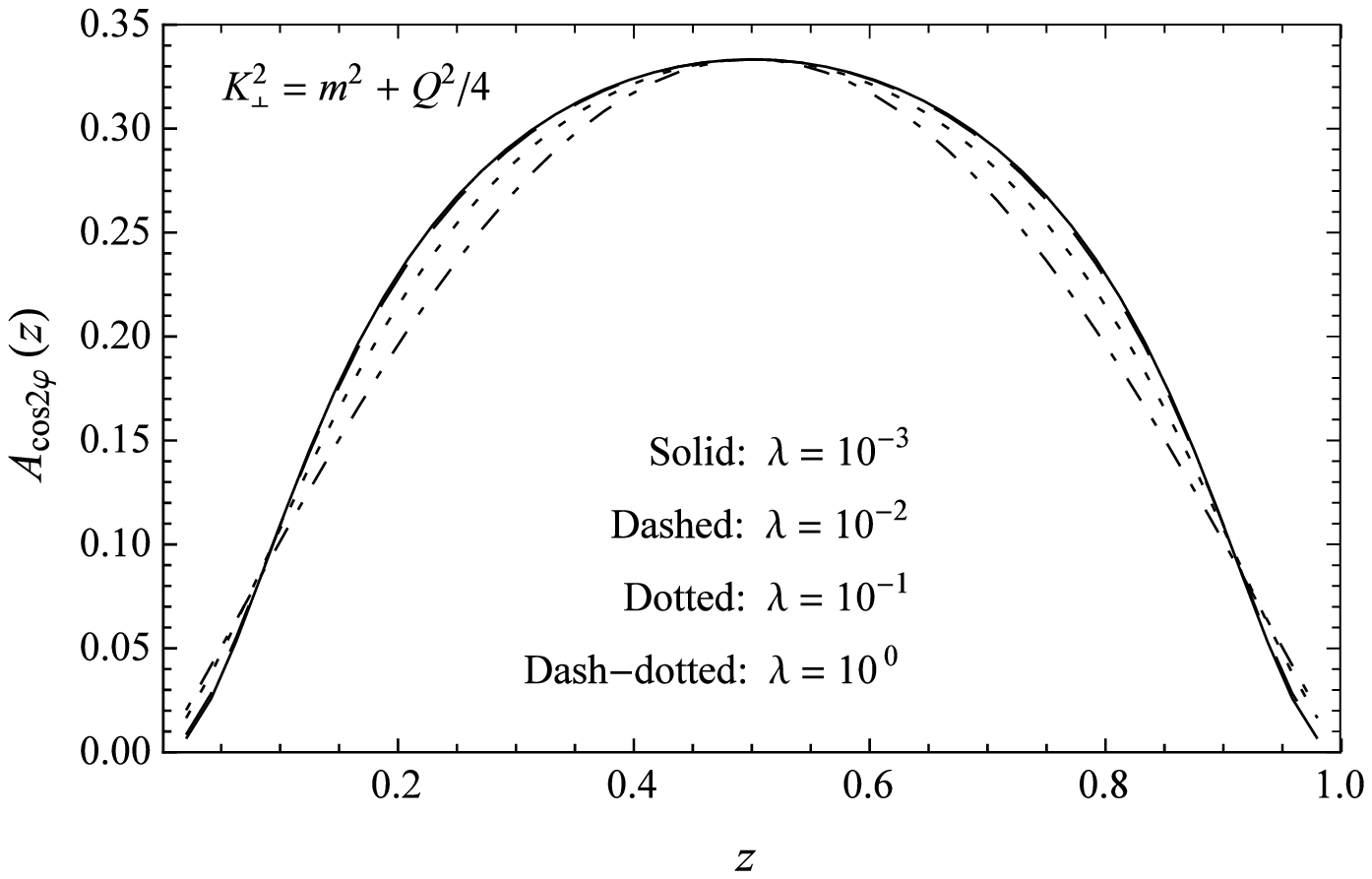}
\caption{\label{Fig1} Azimuthal $\cos2\varphi$ asymmetry $A_{\cos2\varphi}(z)$ in heavy-quark leptoproduction as a function of $z$ at several values of $\lambda=m^2\big/Q^2$.}
\end{minipage}\hspace{1.pc}%
\begin{minipage}{18.2pc}
\includegraphics[width=18pc]{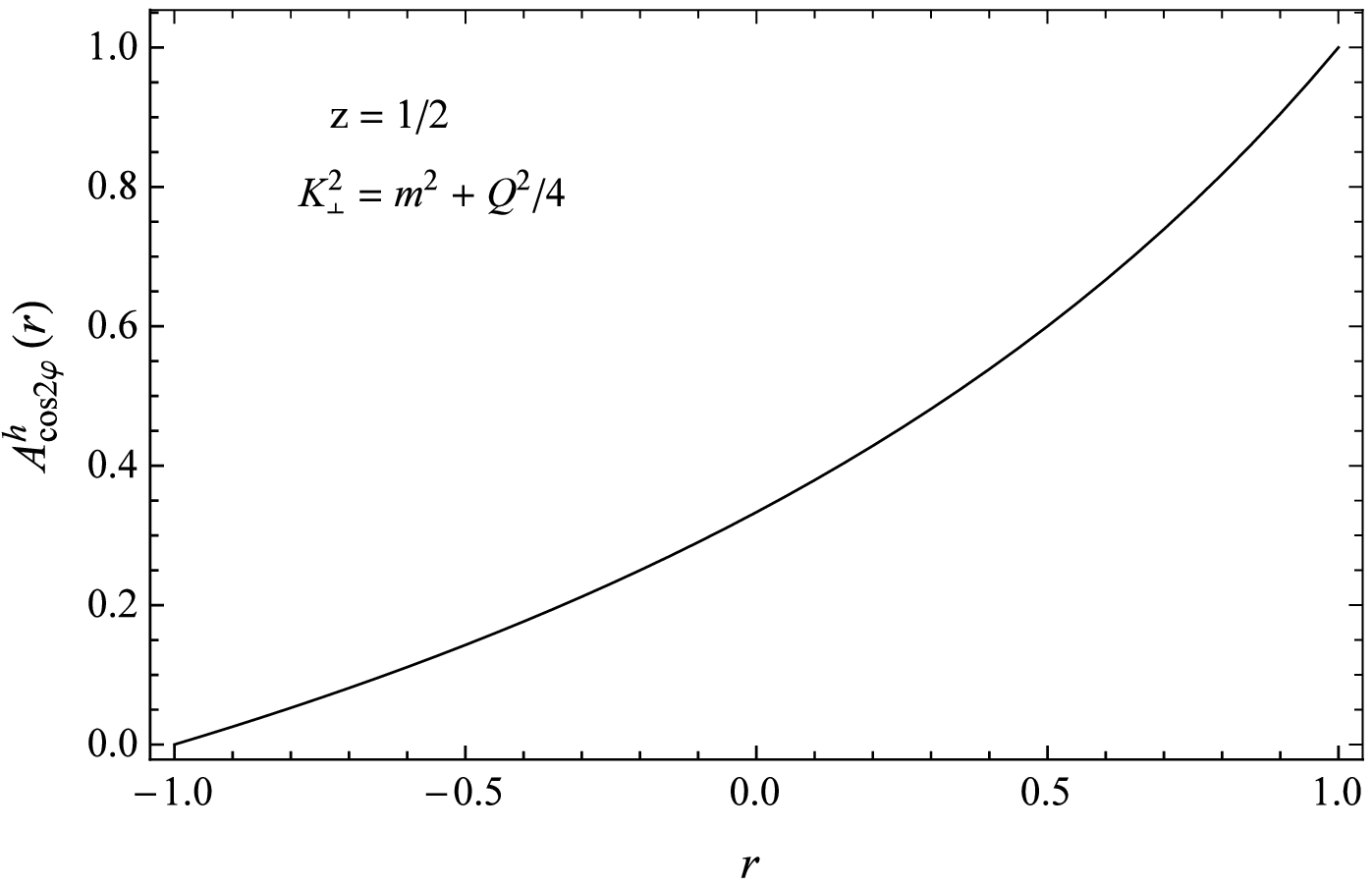}
\caption{\label{Fig2} Maximum value of the $\cos2\varphi$ asymmetry with the contribution of linearly polarized gluons, $A^h_{\cos2\varphi}(r)$, as a function of $r$.}
\end{minipage} 
\end{figure}

At sufficiently small $y\ll 1$ and fixed values of $Q^2$, the $\cos2\varphi$ asymmetry without and with the contribution of linearly polarized gluons ($A_{\cos2\varphi}$ and $ A^h_{\cos2\varphi}$, respectively) is defined as
\begin{align} \label{5}
A_{\cos2\varphi}(z,\vec{K}_{\perp}^2)&\simeq \hat{B}_A\big/\hat{B}_2,& A^h_{\cos2\varphi}(z,\vec{K}_{\perp}^2,r)&\simeq \frac{\hat{B}_A}{\hat{B}_2}\frac{1-2r\hat{B}^h_A\big/\hat{B}_A}{1-2r\hat{B}^h_2\big/\hat{B}_2}.
\end{align}
The function $A_{\cos2\varphi}(z,\vec{K}_{\perp}^2)$ takes its maximal value at $z=1/2$ and $\vec{K}_{\perp}^2=m^2+Q^2/4$, where $m$ is the heavy-quark mass: $A_{\cos2\varphi}(z=1/2,\vec{K}_{\perp}^2=m^2+Q^2/4)=\frac{1}{3}$. The quantity $A_{\cos2\varphi}(z)\equiv A_{\cos2\varphi}(z,K_{\perp}^2=m^2+Q^2/4)$ in heavy-quark leptoproduction as a function of $z$ at several values of $\lambda=m^2\big/Q^2$ is presented in  Fig.~\ref{Fig1}. One can see that this distribution is practically independent of $Q^2$ for both charm and bottom quarks.

The function $A^h_{\cos2\varphi}(z,\vec{K}_{\perp}^2,r)$ has a maximum at $z=1/2$ and $\vec{K}_{\perp}^2=m^2+Q^2/4$ for all values of $r$ in the interval $-1\leq r\leq 1$. As shown in 
Ref.~\cite{we_TMD},
\begin{equation} \label{6}
A^h_{\cos2\varphi}(r)\equiv A^h_{\cos2\varphi}(z=1/2,\vec{K}_{\perp}^2=m^2+Q^2/4,r)=\frac{1+r}{3-r}.
\end{equation}
The function $A^h_{\cos2\varphi}(r)$ is depicted in  Fig.~\ref{Fig2} where its strong dependence on $r$ is seen.

As to the  $\cos\varphi$ asymmetry, it has more complicated behavior. In particular, the quantity $A_{\cos\varphi}(z,\vec{K}_{\perp}^2)$ is an alternating function of both $z$ and $\vec{K}_{\perp}^2$. In detail, this asymmetry is discussed in Refs.~\cite{we_TMD,we_Baldin}.
%\begin{figure}[h]
%\begin{minipage}{14pc}
%\includegraphics[width=14pc]{name.eps}
%\caption{\label{label}Figure caption for first of two sided figures.}
%\end{minipage}\hspace{2pc}%
%\begin{minipage}{14pc}
%\includegraphics[width=14pc]{name.eps}
%\caption{\label{label}Figure caption for second of two sided figures.}
%\end{minipage} 
%\end{figure}
%\begin{figure}[h]
%\includegraphics[width=14pc]{name.eps}\hspace{2pc}%
%\begin{minipage}[b]{14pc}\caption{\label{label}Figure caption for a narrow figure where the caption is put at the side of the figure.}
%\end{minipage}
%\end{figure}
\subsection{Callan-Gross ratio}
\begin{figure}[h]
\begin{minipage}{18.2pc}
\includegraphics[width=18.2pc]{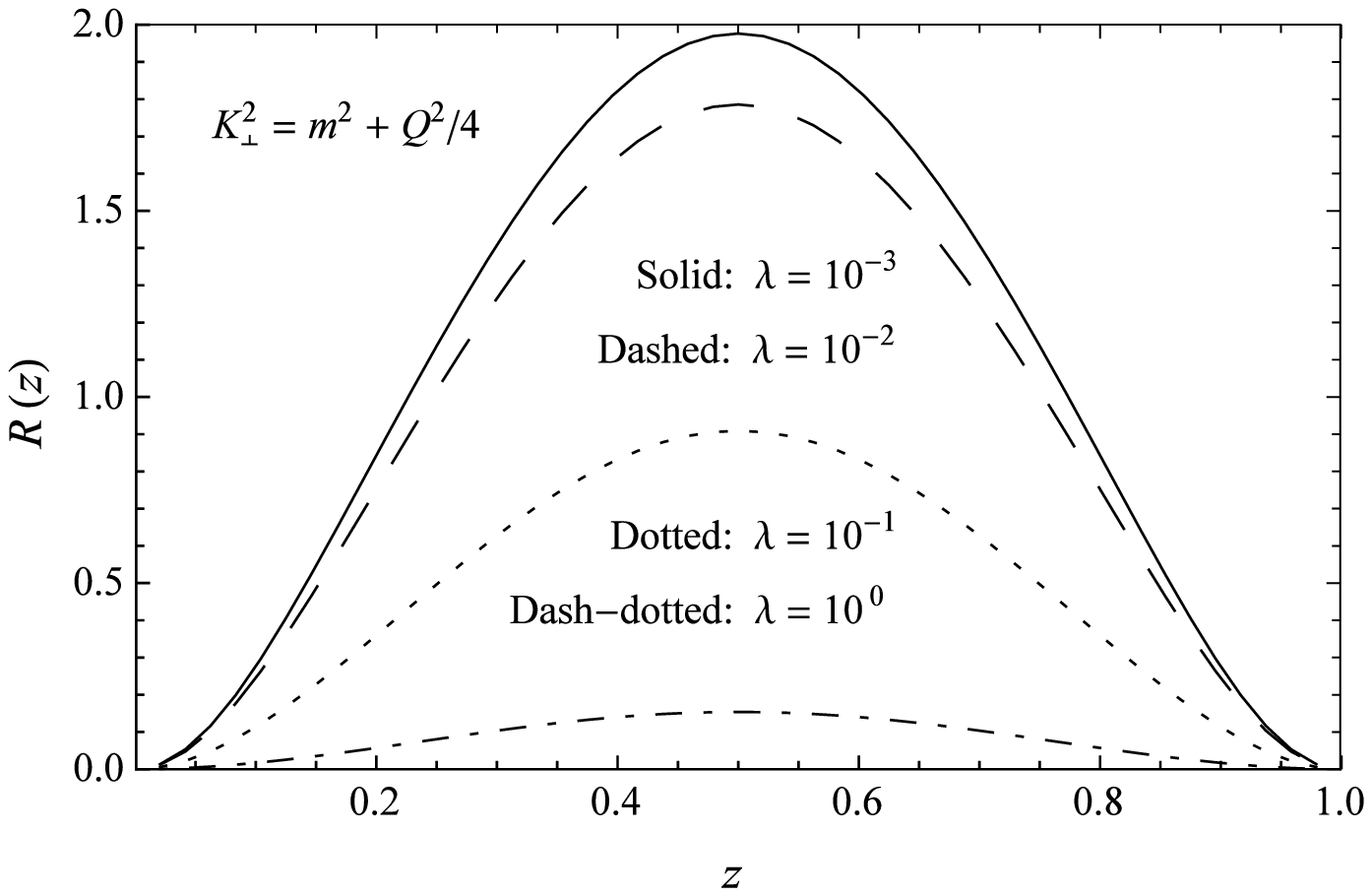}
\caption{\label{Fig3} Callan-Gross ratio $R(z)$ in heavy quark leptoproduction as a function of $z$ at several values of $\lambda$.}
\end{minipage}\hspace{1.pc}%
\begin{minipage}{18.2pc}
\includegraphics[width=18pc]{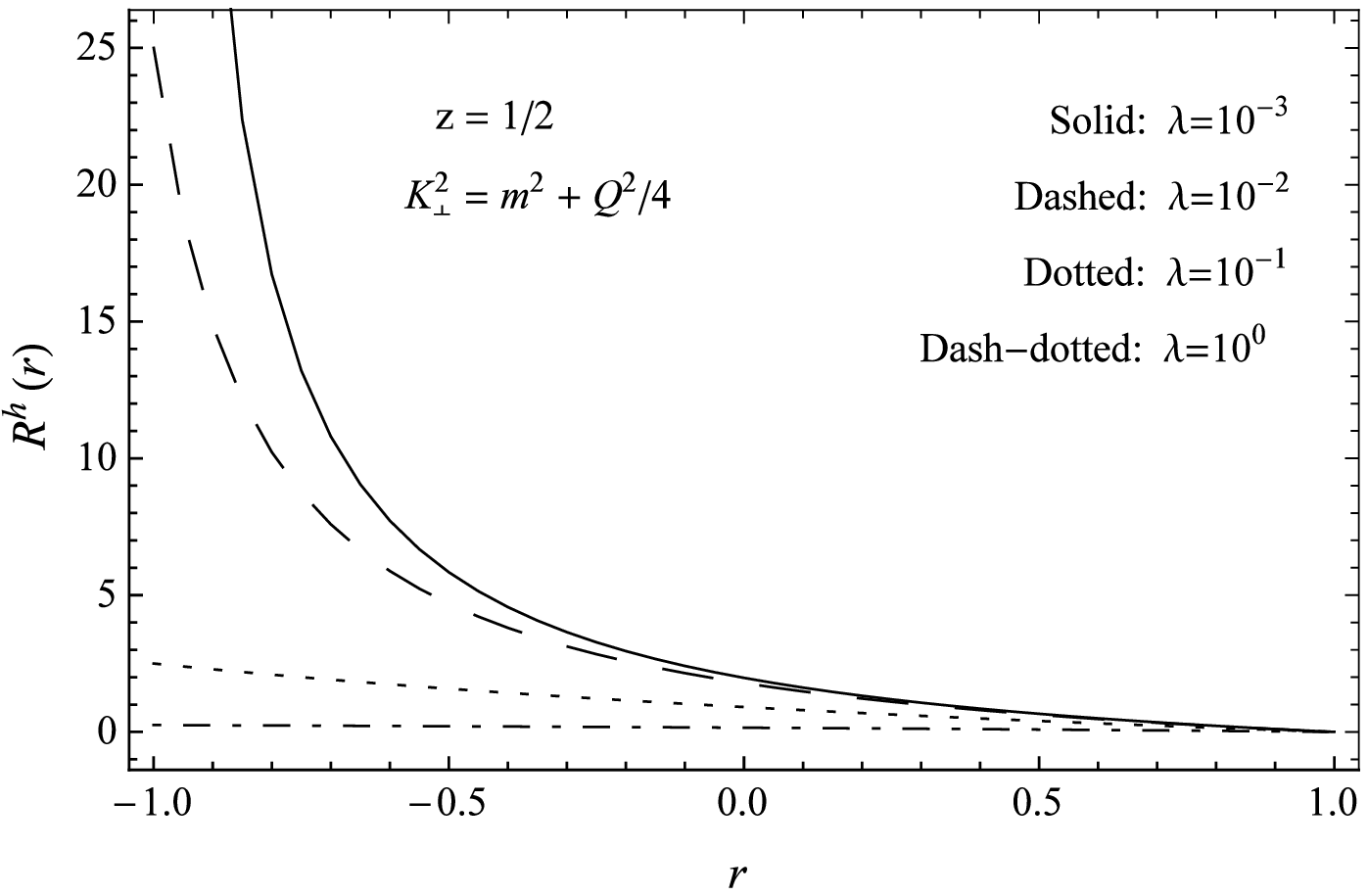}
\caption{\label{Fig4} Maximum value of the Callan-Gross ratio, $R^h(r)$, with the contribution of linearly polarized gluons as a function of $r$.}
\end{minipage} 
\end{figure}

At LO in QCD, the Callan-Gross ratio without and with the contribution of linearly polarized gluons ($R$ and $R^h$, correspondingly) is: 
\begin{align} \label{7}
R(z,\vec{K}_{\perp}^2)&=\hat{B}_L\big/\hat{B}_T ,& R^h(z,\vec{K}_{\perp}^2,r)&=\frac{\hat{B}_L}{\hat{B}_T}\frac{1-2r\hat{B}^h_L\big/\hat{B}_L}{1-2r\hat{B}^h_T\big/\hat{B}_T}.
\end{align}
Like the $\cos2\varphi$ asymmetry, the function $R(z,\vec{K}_{\perp}^2)$ has an extremum at $z=1/2$ and $\vec{K}_{\perp}^2=m^2+Q^2/4$. This maximum value is $R(z=1/2,\vec{K}_{\perp}^2=m^2+Q^2/4)=\frac{2}{1+12\lambda}$. This fact implies that the maximum value of the ratio $R$ at high $Q^2\gg m^2$ (i.e. for $\lambda\rightarrow 0$) is large, about 2. Fig.~\ref{Fig3} shows the Callan–Gross ratio $R(z)\equiv R(z,\vec{K}_{\perp}^2=m^2+Q^2/4)$ in heavy quark leptoproduction as a function of $z$ at several values of $\lambda$.

The function $R^h(z,\vec{K}_{\perp}^2,r)$ has also a maximum at $z=1/2$ and $\vec{K}_{\perp}^2=m^2+Q^2/4$ for all values of $r$ in the interval $-1\leq r\leq 1$. As shown in Ref.~\cite{we_TMD_CG}, 
\begin{equation} \label{8}
R^h(r)\equiv R^h(z=1/2,\vec{K}_{\perp}^2=m^2+Q^2/4,r)=\frac{2(1-r)}{1+r+4\lambda\,(3-r)}.
\end{equation}
The quantity $R^h(r)$ is depicted in Fig.~\ref{Fig4}. One can see that $R^h(r)$ vanishes for $r=1$. Note also an unlimited growth of the Callan-Gross ratio with $Q^2$ at $r\rightarrow-1$, $R^h(r=-1)=\frac{Q^2}{4m^2}$. 
\section{Conclusion}
Physically, our main observation can be formulated as follows. When a linearly polarized gluon interacts with transverse virtual photon, the heavy-quark production plane is preferably  orthogonal to the direction of the gluon polarization. On the contrary, the longitudinal component of the cross section $g^{\uparrow}\gamma^{\,*}\rightarrow Q\bar{Q}$ takes its maximum value when the momenta of emitted quarks and the gluon polarization lie in the same plane.

We conclude that the azimuthal asymmetries and Callan-Gross ratio in heavy-quark pair  leptoproduction could be good probe of the linear polarization of gluons inside unpolarized proton. 

{\bf Acknowledgements.} We are grateful to Ter-Antonyan-Smorodinsky program of BLTP (JINR) for financial support.

\end{document}